\documentclass[acmsmall,manuscript]{acmart} % [from CHI 2021] https://chi2021.acm.org/for-authors/chi-publication-formats
\renewcommand\footnotetextcopyrightpermission[1]{} % removes footnote with conference information in first column
\usepackage{booktabs}
\usepackage{graphicx}
\usepackage{listings}
\usepackage{xcolor}
\usepackage{array}
\usepackage{ragged2e}
\usepackage{hhline}
\usepackage{colortbl}
\usepackage{todonotes}
\usepackage{svg}
\usepackage{amsmath}
\usepackage{enumitem}
\usepackage{setspace}
\usepackage{float}

\AtBeginDocument{%
  \providecommand\BibTeX{{%
    \normalfont B\kern-0.5em{\scshape i\kern-0.25em b}\kern-0.8em\TeX}}}

\begin{document}

\title[XAI Reloaded] {Explainable AI Reloaded: Challenging the XAI Status Quo in the Era of Large Language Models}

\author{Upol Ehsan}
\affiliation{%
  \institution{Georgia Institute of Technology}
  \country{USA}}

 \author{Mark O. Riedl}
\affiliation{%
  \institution{Georgia Institute of Technology}
  \country{USA}}

\renewcommand{\shortauthors}{Ehsan \& Riedl}

\begin{abstract}
% NEW ABSTRACT
When the initial vision of Explainable (XAI) was articulated, the most popular framing was to open the (proverbial) “black-box” of AI so that we could understand the inner workings. With the advent of Large Language Models (LLMs), the very ability to open the black-box is increasingly limited especially when it comes to non-AI expert end-users. In this paper, we challenge the assumption of “opening” the black-box in the LLM era and argue for a shift in our XAI expectations. Highlighting the epistemic blind spots of an algorithm-centered XAI view, we argue that a human-centered perspective can be a path forward. We operationalize the argument by synthesizing XAI research along three dimensions: explainability outside the black-box, explainability around the edges of the black box, and explainability that leverages infrastructural seams. We conclude with takeaways that reflexively inform XAI as a domain.

\end{abstract}


% \ccsdesc[500]{Human-centered computing~User studies}
% \ccsdesc[300]{Human-centered computing~Empirical studies in collaborative and social computing}


\maketitle

\section{Provocation}

\textit{With the advent of Foundation Models \& Large Language Models like ChatGPT, is “opening the black-box” still a reasonable and achievable goal for Explainable AI (XAI)? Do we need to shift our perspectives? }

In February of 2023, Nadeem (pseudonym), a relative of the first author of this article, asked if ChatGPT could be used to do homework. Nadeem is a high-schooler who shared that he was always under a tight deadline and needed to be more “efficient” with his homework. He heard from his friends that ChatGPT can help summarize papers or books, which can make Nadeem more “productive” with his homework. 

\begin{figure*}[tbh]
    \vspace{-2mm}
    \centering
    \includegraphics[width=0.7\columnwidth]{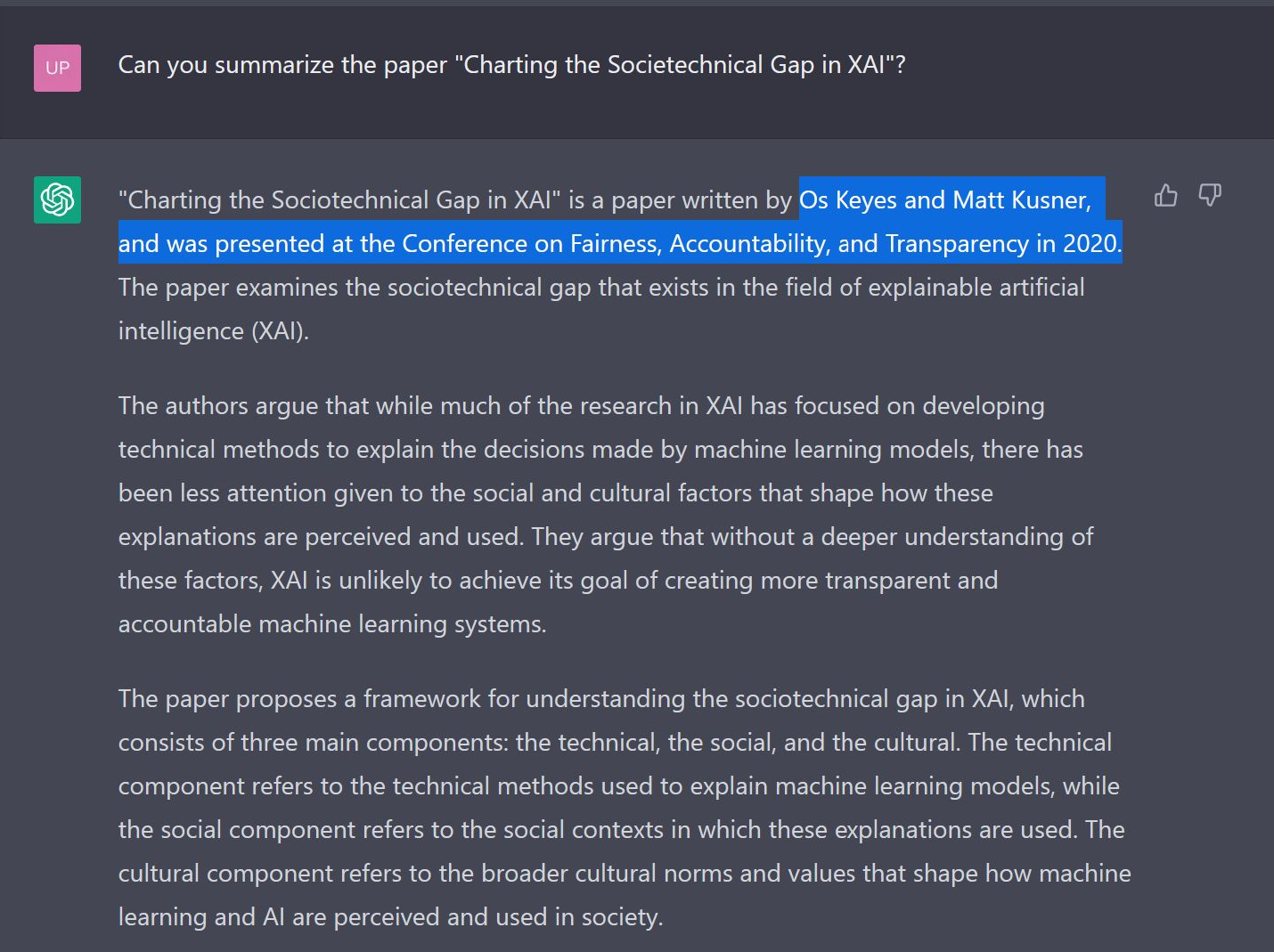}
    % \begin{fignote}
    % Note: 
    % \end{fignote}
    \caption{Screenshot of ChatGPT’s summary of a paper with hallucinated or confabulated content.}
    \label{fig:chatGPT}
    \Description[Screenshot from ChatGPT’s response to summarizing a paper showcasing its hallucinated or confabulated content]{Figure 1: Screenshot from ChatGPT reads:
    Prompt: Can you summarize the paper (redacted)?
    Response: (redacted) is a paper written by Os Keyes and Matt Kusner, and was presented at the Conference on Fairness, Accountability, and Transparency in 2020. The paper examines the sociotechnical gap that exists in the field of explainable artificial intelligence (XAI).
    The authors argue that while much of the research in XAI has focused on developing technical methods to explain the decisions made by machine learning models, there has been less attention given to the social and cultural factors that shape how these explanations are perceived and used. They argue that without a deeper understanding of these factors, XAI is unlikely to achieve its goal of creating more transparent and accountable machine learning systems.
    The paper proposes a framework for understanding the sociotechnical gap in XAI, which consists of three main components: the technical, the social, and the cultural. The technical component refers to the technical methods used to explain machine learning models, while the social component refers to the social contexts in which these explanations are used. The cultural component refers to the broader cultural norms and values that shape how machine learning and Al are perceived and used in society.}
\end{figure*}

Before responding to Nadeem, ChatGPT was taken for a test drive. It was prompted to summarize an academic paper (Figure~\ref{fig:chatGPT} similar to how Nadeem might use it – as someone who was not an AI researcher or experienced prompting Large Language Models (LLMs). Fortunately, ChatGPT generated a coherent response. ChatGPT gave the names of the authors of the paper and details about the paper’s publication at ACM FAccT 2020. Unfortunately, the names of the authors and publication details were made up! The confabulated author names were immediately obvious because we wrote the paper that was prompted to be summarized~\cite{ehsan2023charting}. However, the rest of the details was extremely plausible – the paper very well could have appeared at that conference, but did not. The first author of this paper almost missed another detail in ChatGPT’s summary. The original paper described a framework with two dimensions: social and technical. The generated summary claimed the framework described three dimensions: social, technical, and cultural, which, while wrong, was plausible enough that even the very author of the paper almost missed that crucial inaccuracy!

\vspace{-7pt}
\subsection{Separating Fact from Fiction}
The true story above demonstrates the effortful process required to disentangle fact from fiction in GPT’s output, even from someone knowledgeable of the source material. Even more notably, there was no way for our protagonist, an expert in Explainable AI, to “open” the black-box of ChatGPT and understand why it produced what it produced or where it might be faithful to the facts or prone to confabulation (also called hallucination). On the one hand, he lacked access to the internal details such as the parameters of the model. On the other hand, even if one did have access to the internal parameters of the model, given the scale and complexity of the neural architecture of such a large language model, interpreting it is unlikely to produce human-understandable and actionable information. 

\vspace{-7pt}
\section{Tensions: XAI and Large Language Models}
The field of Explainable AI (XAI) is concerned with developing techniques, concepts, and processes that can help stakeholders understand the reasons behind the AI system’s decision-making~\cite{liao2021human,ehsan2019automated}.

For our purposes, we adopt a design lens in XAI that is sociotechnically-informed~\cite{liao2021human,ehsan2020human,dhanorkar_who_2021} and adopt the broad definition that an explanation is an answer to a \textit{why}-question~\cite{miller2019explanation,dennett1989intentional,lewis1986causal}. Given AI systems exist in sociotechnical settings~\cite{sun2022investigating, liao2021question}, it takes more than just algorithmic transparency to make them explainable~\cite{ehsan2022human,miller2019explanation}. 
Thus, explaining what is happening “inside the black box” often requires us to also understand things “outside the black box” ~\cite{ehsan2021expanding, dhanorkar_who_2021, liao2020questioning}, requiring us to consider the entire AI lifecycle (vs. just the algorithm). 
For instance, why a facial recognition system disproportionately misclassified women of color~\cite{buolamwini2018gender} can be explained by looking at demographic compositions in the training data.
A sociotechnically situated view of XAI expands the concept of explainability beyond the bounds of the algorithm~\cite{ehsan2021expanding} and positions it as a relational and audience-dependent construct instead of a model-inherent one~\cite{arrieta2020explainable,mohseni2018multidisciplinary,arya2019one,miller2019explanation}. 
Emerging work ~\cite{schoeffer2021appropriate, haque2020understanding, pushkarna2022data} showcases how a broader XAI perspective can potentially address criticisms of popular algorithm-centered XAI techniques, which can be ineffective~\cite{alqaraawi2020evaluating,poursabzi2018manipulating,zhang2020effect} and potentially risky~\cite{kaur2020interpreting,stumpf2016explanations}.

When we consider a service such as ChatGPT, GPT-4, Microsoft Copilot, Google Gemini, Claude, or Meta AI, what prospects are there for ``opening'' the black-box of AI? These models have hundreds of billions of parameters, all acting in conjunction to generate a distribution over possible words to choose from to build a response, word by word. If we had access to all the weights, could we interpret and explain the model? If we had access to the parameters of a model and the activation values for an input could we interpret and explain the model? In the case of the above large language models the point is moot. All these models run on servers behind APIs that do not allow inspection of the neuron activations and weights. However, even if we could access this information, the raw values of weights and activations are meaningless to most people without synthesizing some visualization or text summarization that provides a lay-understandable analysis of the internal operations of the system and how the results were generated by the system. Consider OpenAI’s work on interpreting the patterns that cause individual neurons to activate~\cite{bills2023language}. How would knowing what causes neuron \#2142 to activate have helped Nadeem, a non-AI expert, know how to better use ChatGPT to complete his homework? What actionable information from this neural activation pattern can a non-AI expert use meaningfully?

LLMs are increasingly being incorporated as components in systems that chain multiple processes together. 
In particular, Retrieval Augmented Generation (RAG) combines LLMs with web search such that a web retrieval module first retrieves relevant documents, which are then used to inform an LLM~\cite{lewis2020retrieval}.  
While opening the black-box LLMs may not yield actionable explanations, modular architectures afford the ability to inspect and explain how data is changed going in and out of black-box modules.

\section{Is Explainable AI doomed to fail?}
Despite the commendable progress in algorithm-centered approaches in XAI, there are significant deficiencies. Studies examining how people actually interact with AI explanations have found popular XAI techniques to be ineffective~\cite{alqaraawi2020evaluating, poursabzi2018manipulating,zhang2020effect}, potentially risky~\cite{kaur2020interpreting,stumpf2016explanations}, and even obsolete in real-world deployed contexts~\cite{liao2020questioning}. XAI developers tend to design explanations \textit{as if} people like them are going to use their systems, earning an infamous reputation of “inmates running the asylum” ~\cite{miller2019explanation}. 
In fact, a majority of current deployments serve AI engineers instead of end-users whose needs are ignored~\cite{bhatt2020explainable}. This creates a gap between design expectations and reality— how developers envision the designed AI explanations to get interpreted and how users actually perceive those explanations in reality.

As Large Language Models (LLMs) become prominent, is Explainable AI – a research area in flux and its infancy – doomed to fail? No. There is hope. Before we throw in the towel, there are a few things to consider. 

\subsection{AI systems are Human-AI assemblages}
First, the techno-centric, algorithm-centered, discourse of XAI fails to appreciate the sociotechnical reality of AI systems. When we say “AI systems,” what we very often mean to say is “Human-AI assemblages,” where the “human” part of the Human-AI assemblage is often implicit~\cite{ehsan2021expanding}. No real-world AI systems work in a vacuum. Black-boxes \textit{by themselves} do not do the work – humans \textit{with} black-boxes do the work~\cite{ehsan2020human}. Even if the human contribution to the work is to just provide an input, this is a significant contribution because AI systems are useful to people as tools. Thus, the explainability of AI systems entails explainability of the Human-AI assemblage, which has at least two components: the human (or humans) and the AI~\cite{ehsan2021expanding,ehsan2023charting}. Thus, how can we achieve the explainability of the Human-AI assemblage by just focusing on the explainability of the AI model? \textit{XAI is therefore not just technical, it is sociotechnical}. It requires more than just algorithmic transparency – more than being able to open the black box. 
% \vspace{-4mm}
\begin{figure*}[tbh]
    \centering
    \includegraphics[width=0.8\columnwidth]{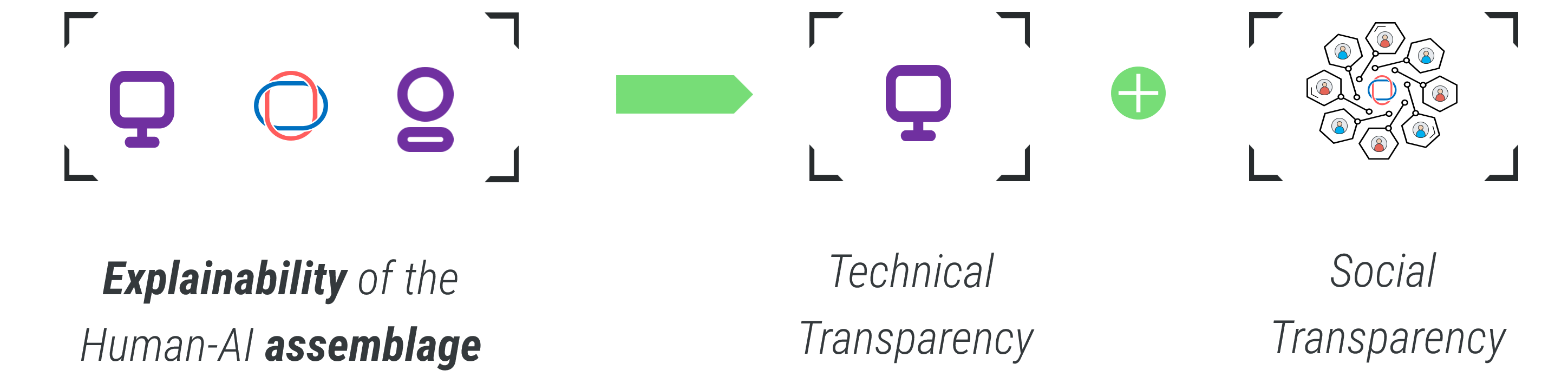}
    % \begin{fignote}
    % Note: 
    
    % \end{fignote}
    \caption{Illustrating how the explainability of the Human-AI assemblage is more than just technical (algorithmic) transparency}
    \label{fig:human-AI assemblage}
    \Description[Illustrating how the explainability of the Human-AI assemblage is more than just technical (algorithmic) transparency]{Figure 2: From left to right, diagram starts a picture of Human-AI assemblage with icons of a computer and human followed by an arrow that splits into technical transparency and social transparency showcasing how explainability of the assemblage includes things beyond algorithmic/technical transparency} 
\end{figure*}

Second, what we mean by ``AI'' is evolving. Compared to AI systems even five years ago, the Deep Learning systems in the Foundation Model era, such as LLMs, are much more complex, have orders of magnitude more parameters, and are running at unprecedented scales. Thus, AI as a \textit{design material} is tricky and is evolving~\cite{yang2020re,dove_ux_2017,ehsan2023charting}. Our understanding and expectations of what it means for AI-as-design-material to be explainable should also evolve. Further, XAI techniques that focus solely on the algorithm or the model face a new challenge: it is getting increasingly hard to open the black box! As AI systems are increasingly end-user facing, those that need the explanations the most are on the other side of an AI or user interface. This is the case for the most popular Large Language Models and chatbots, and it is also the case for other types of consumer-facing systems. When the initial vision of XAI was articulated, a popular framing was to “open” the (proverbial) “black-box” of AI~\cite{castelvecchi2016can,nott2017explainable}, so that we could see inside of it, figure out what it was doing, why it was doing it, and if it was doing it correctly. With the advent of large language models, that ability to open the black-box is increasingly limited due to the sheer complexity of the models and the increased prevalence of models behind restrictive APIs. And even if we did manage to “open” it, we will not understand what we see. 

\section{Human-Centered Explainable AI: Beyond Algorithmic Transparency}

Given AI systems are bounded by their training data, by construction, they cannot incorporate the real-world dynamics "outside" the black-box. Thus, an algorithm-centered view of XAI is--by construction--a limiting view, one that handicaps the XAI system from doing what we want to do-- solve real world problems. We need a paradigm that can accommodate an expansion of the epistemic canvas-- an increase of the aperture of the viewing lens-- to include the sociotechnical dynamics in which XAI systems are embedded so that we can do what we set out to do -- solve real world problems. 

This is where the domain of \textit{Human-Centered Explainable AI (HCXAI)}~\cite{ehsan2020human} can help. 
HCXAI is a holistic vision of AI explainability, one that is human-centered and sociotechnical in nature. Situated as a Critical Technical Practice~\cite{agre1997toward,agre1997computation}, it draws its conceptual DNA from critical AI studies and HCI (e.g., reflective design~\cite{sengers2005reflective, dourish2004action,dourish2004reflective}, value-sensitive design~\cite{friedman2008value}).
HCXAI encourages us to critically reflect and question dominant assumptions and practices of a field, such as algorithm-centered XAI. It also adopts a value-sensitive approach to both users and designers in the development of technology that challenges the status quo of a field. HCXAI encapsulates the philosophy that not everything that is important lies inside the black box of AI. Critical insights can lie outside it. Why? \textit{Because that’s where the humans are.}

Thinking \textit{outside} the black box of AI can help us meet our goals of helping people understand and calibrate their trust in AI systems. Even if we cannot meaningfully open the black box or interpret its complexities, there are a lot of things we can do to understand and explain the system \textit{around} the black box. Increasing the aperture of XAI can help us focus on the most important part: \textit{who} the human(s) is (are), what they are trying to achieve in seeking an explanation, and how to design XAI techniques that meet those needs. Indeed, explanations of the sociotechnical system can offer us an important affordance: \textit{actionability}~\cite{singh2023directive,joshi2019towards,ehsan2021explainable}.

At its core, actionability is about what a user can do with the information in an explanation~\cite{ehsan2021explainable}. An actionable XAI system empowers the user by increasing the space of possible informed actions to achieve their end goals. This could be understanding how to change the inputs, contesting a decision, or learning when and how to use the system more appropriately. Actionability also addresses another important question: how do we know if an XAI system is useful? There are an increasing number of reports of XAI systems that are deployed and fail to have any measurable impact on their users~\cite{alqaraawi2020evaluating,stumpf2016explanations}. Many of these systems failed because the XAI systems were not designed with user needs in mind, such as by providing users with information they could already intuit themselves, by providing information that was onerous to verify, or by providing information that users could not use. In other words, the explanations generated by the systems were not actionable. 

\section{The way forward}
With the reframing around human-AI assemblages and XAI systems that place the human as the central concern, and armed with actionability as the metric for success, we now lay out three possible paths forward. This list is not meant to be exhaustive or prescriptive. It is meant to be generative by providing emerging evidence for how Human-Centered XAI (HCXAI) can address the growing needs for understanding our increasingly AI-infused world.

\subsection{Explainability outside the black-box: Social Transparency}

Most consequential AI systems are embedded in organizational environments where groups of humans interact with it. These real-world AI systems, as well as the explanations they produce, are socially-situated~\cite{liao2020questioning,ehsan2021operationalizing}. Therefore, the socio-organizational context in which these systems are used is key. Why are we not incorporating socio-organizational contexts into how we think about explainability in AI? How can we tackle the explainability of Human-AI assemblages?

Enter Social Transparency (ST) a sociotechnically-informed perspective that incorporates the socio-organizational context into explaining AI-mediated decision-making~\cite{ehsan2021expanding}. Social transparency allows us to augment the explainability of a human-AI assemblage without necessarily changing anything about the AI model. Social transparency allows one to annotate an output or behavior from an AI system with the 4W \textbf{who} did \textbf{what}, \textbf{when} and \textbf{why}. These annotations are shared between others using the system. They allow users to see whether and why others have accepted or rejected an AI’s output. Social transparency does two important things: first, it challenges the dominant narrative of algorithm-centered notions of XAI; second, it expands our understanding of XAI beyond technical transparency by illustrating how adding social context can help people make better, more actionable decisions with AI systems.

\begin{figure*}[tbh]
    \centering
    \includegraphics[width=0.85\columnwidth]{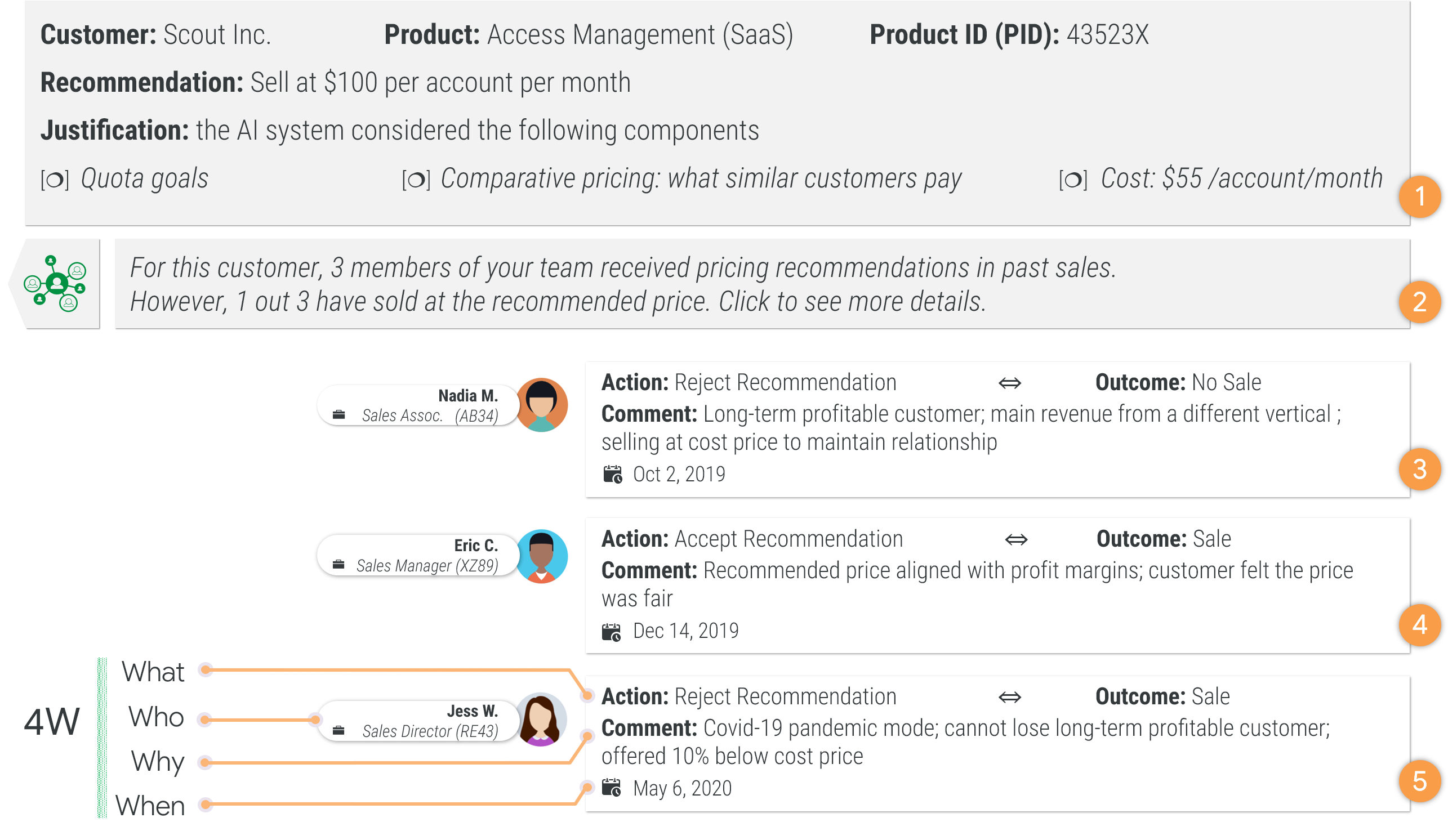}
    \caption{Sales scenario with Social Transparency (ST) used in~\cite{ehsan2021expanding} (reproduced with permission from authors). The labeled blocks are: (1) Decision information and model explanation: Information of the current sales decision, the AI's recommended price and a ``feature importance'' explanation justifying the model's recommendation, inspired by real-world pricing tools; (2) ST summary: Beginning of ST giving a high-level summary of how many teammates in the past had received the recommendation and how many sold at the recommended price; (3-5): ST blocks with "4W" features containing the historical decision trajectory of three other users.}
    \label{fig:ST_visualScenario}
    \Description[Figure showing the anatomy of the visual scenario (sales context).]{Figure 3: Visual scenario by labeled blocks: (1) Decision information and model explanation: Information of the current sales decision, the AI's recommended price and a ``feature importance'' explanation justifying the model's recommendation, inspired by real-world pricing tools; (2) ST summary: Beginning of ST giving a high-level summary of how many teammates in the past had received the recommendation and how many sold at the recommended price; (3-5): ST blocks with "4W" features containing the historical decision trajectory of from each of the colleagues.}
    \vspace{-10pt}
\end{figure*}

Imagine the following scenario (Figure~\ref{fig:ST_visualScenario}): Aziz is a software seller trying to use a powerful AI-based pricing tool to do something consequential: offer the right price to a client company. The AI suggests a price. Moreover, its suggestion has technical transparency – it explains its recommendation by showing Aziz the top features it considered, such as sales quota goals, comparative pricing with other clients, and costs. Confident with the AI’s recommendation, Aziz makes a bid, but the client finds the price too high and walks out. 

Despite an accurate AI model and the presence of technical transparency, why did the bid fail? There could be algorithmic reasons for it. But might also be relevant contextual factors outside the box that can help explain why the bid failed. Perhaps the history between Aziz and the client that was not honored? Or maybe there were external events that happened since the model was trained, such as a pandemic-induced budgetary crisis. 

Now imagine that Aziz could see that more than 65\% of his peers rejected the AI’s pricing recommendation (Block 2 in Fig.~\ref{fig:ST_visualScenario}). Or, what if Aziz knew that Jess, a director in the company, sold the product at a loss due to pandemic-related budgetary cuts?(Block 5 in Fig.~\ref{fig:ST_visualScenario})

This peripheral vision of \textit{who did what, when and why} – called the \textbf{4W} – are the constitutive design elements of Social Transparency that can encode relevant socio-organizational context. The benefit of taking a holistic approach to explainability is clear: a study of real-world AI users in sales, cybersecurity, and healthcare found that social transparency, in the form of the 4W, helped people calibrate their trust in the AI’s performance, provide actionable information for AI contestability and robust decision-making, and the organizational context made visible enabled better collective actions in the organization and strengthened the human-AI assemblages~\cite{ehsan2021expanding}.

By incorporating the socio-organizational context, Social Transparency makes our understanding of XAI more holistic, representing the Human-AI assemblage more realistically than a purely algorithm-centered XAI view. We should note that Social Transparency is agnostic to whether an AI system is black-boxed or not. As long as there is an AI-based recommendation or decision, we can attach 4W – the socio-organizational context – to it. In a completely black-boxed AI system, there will not be any technical transparency. However, the 4W can add transparency to the social side of the Human-AI assemblage.

\subsection{Explainability around the Edges of the Black Box: Rationale Generation \& Scrutability}

If the black box cannot be cracked open in any meaningful sense, there is another possibility: incorporate explainability \textit{around the edges} of the black box to foster a better functional understanding in the user~\cite{paez2019pragmatic} such that it fosters actionability. One of the original formulations of rationale generation~\cite{ehsan2019automated} postulated that there was no need to know how a black box worked as long as we could learn how to give actionable advice about the black box by looking at its inputs and outputs. It was philosophically grounded in Fodor’s work on Language of Thought~\cite{fodor1975language}: how is it that, despite not having a 1-1 neural correlate of thought, humans can effectively communicate by translating their thoughts into words? For Human-AI interaction, even if the exact mechanism of the (artificial) neural correlate of AI’s thought was not known to the human, as long as actionable information is present in the explanation from an AI agent, the Human-AI interaction can proceed. In short, explanations that do not directly access the model can still generate actionable information. 

In the case of large language models, the actionable information is whether any particular input is likely to produce a reliable response that can be trusted. Large language models might be generally capable at many tasks such as question-answering, they are not infallible, and it is always possible for a user to ask a question that results in a confabulation (also called a “hallucination”) that the user is unable to vet. In this case, we can directly use the API to probe how it responds to particular stimuli~\cite{xie2022calibrating}. It is proposed that an XAI system can decompose the original, human authored question into a series of more fine-grained, related questions that provide more opportunities for the model to confabulate responses if it is not competent at the original question. These sub-questions can be selected to be easier for the user to vet. Generating questions to challenge an LLM has been demonstrated to increase users’ ability to determine whether the answer should be trusted or not.

\subsection{Explainability by Leveraging Infrastructural Seams: Seamful XAI}
No AI system is perfect. Mistakes are inevitable. Breakdowns in AI systems often occur when the assumptions we make in design and development do not hold true when they are deployed in the real-world. For example, an AI system can fail when it is trained on data from North America but deployed in South Asia, especially when the end user is unaware of this infrastructural mismatch. These mismatches between design assumptions and real-world usage are called \textit{seams}~\cite{ehsan2022seamful}. Handling the mistakes from AI systems is hard, especially when the AI’s decision-making is hidden or black-boxed. Although black-boxing AI systems can make the user experience \textit{seamless} and easy to use, concealing the seams can lead to downstream harms for end-users, such as uncritical AI acceptance. What can we do differently? How do we move beyond seamless AI? And what can we gain by doing so?

Seamful XAI is a design lens that incorporates the principles of seamful design~\cite{chalmers2003seamful} to augment explainability and user agency. A classic example of seamful design is a "seamful map" of WiFi coverage in your home. If you know the WiFi’s dead zones in your home, you will be able to best use it because you can then avoid.Without revealing the seams, users can have reasonable expectations of perfect WiFi. The map makes the seams in the WiFi’s infrastructure visible to users, which allows them to recalibrate their expectations and behavior. A seamful design principle asks us to leverage the weakness in opportunistic ways~\cite{gaver2003ambiguity}.

Unlike seamlessness, \textit{seamful design does not aim to hide the infrastructure}. Rather, it puts the infrastructure and all its imperfections front and center. Seamful design helps us recognize and grapple with the complex infrastructures systems reside in. Conversely, seamless design ideals risks making the labor it takes to make the system work invisible (e.g., datawork, ghostwork, maintenance work). And, as invisible work is invariably unaccounted for and unappreciated, workers who conduct this work will feel undervalued or invisible. Seamfulness embraces the imperfect reality of spaces we inhabit and makes the most out of it. 

In the context of AI, \textit{seams can be conceptualized as mismatches, gaps, or cracks in assumptions between the world of how AI systems are designed and the world of how AI systems are used in practice}. Seamful XAI seeks to empower users with information that augments their agency by identifying gaps between ideal design assumptions and reality. 

At the heart of Seamful XAI are four observations: 
\begin{enumerate}
    \vspace{-5pt}
    \item Seams are inevitable, arising from the integration of heterogeneous sociotechnical components during technology deployments.
    \item Seams are revealed through system breakdowns.
    \item Instead of treating seams as problematic negatives to be erased, they can be used strategically to calibrate users' reliance and understanding of an AI system.
    \item The goal of this strategic revelation (and concealment) is to support user agency (actionability, contestability, and appropriation).
\end{enumerate}

\textbf{Seamful XAI Design Process: }Let’s review the design process proposed by~\cite{ehsan2022seamful}. 

\begin{figure*}[H]
    \centering
    \includegraphics[width=0.95\columnwidth]{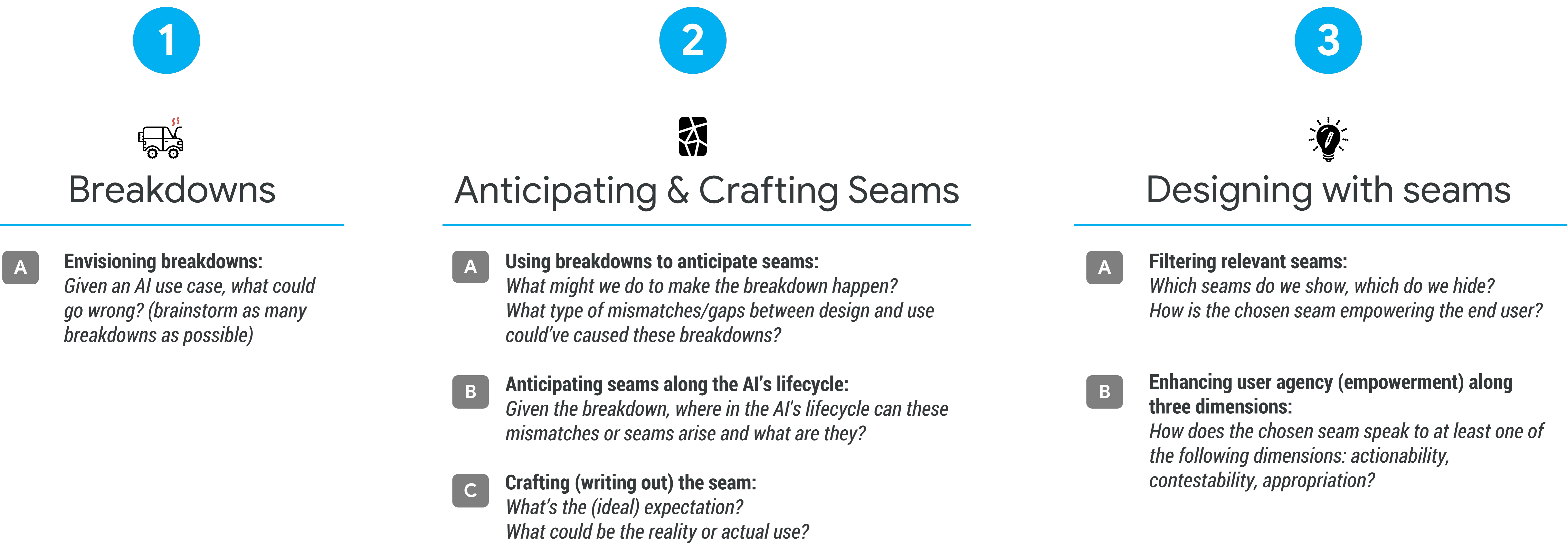}
    % \vspace{-1em}
    \caption{An overview of the Seamful XAI design process used in~\cite{ehsan2022seamful} (reproduced with permission).} 
    \vspace{-1.5em}
    % \textit{more caption items to follow}}
    % \end{fignote}
            \label{fig:SXAI_process123steps}
    % \Description[TBD]{TDB}
    \Description[figure]{Figure 4: An overview of the Seamful XAI design process with key questions relevant to each step. The image has three columns with different line items under each of them. First, breakdowns which has one procedural item - that of "Envisioning Breakdowns" - given an AI use case, what can go wrong? Second, anticipating and crafting seams that has three procedural items - that of using breakdowns to anticipate seams (e.g., what can we do to make the breakdown happen?), of anticipating seams along the AI's lifecycle (e.g., given a breakdown, where in the lifecycle can it emerge from?), and of crafting the seams (e.g., what is the ideal expectation and what could be reality?). Third, designing with seams which has two procedural items - that of filtering relevant seams (e.g., which seams to show or hide?) and enhancing user agency (i.e., how does the chosen seam speak to at least one of the following dimensions - actionability, contestability, and appropriation?).}
\end{figure*}

The \textit{first} step of the process begins with generating "breakdowns." Breakdowns are answers to the question, "what could go wrong when this technology gets deployed?" Answers could include technology failures, unfair treatment of groups, inducing over-reliance, or deskilling. 

\begin{figure*}[t]
    \centering
    \includegraphics[width=0.95\columnwidth]{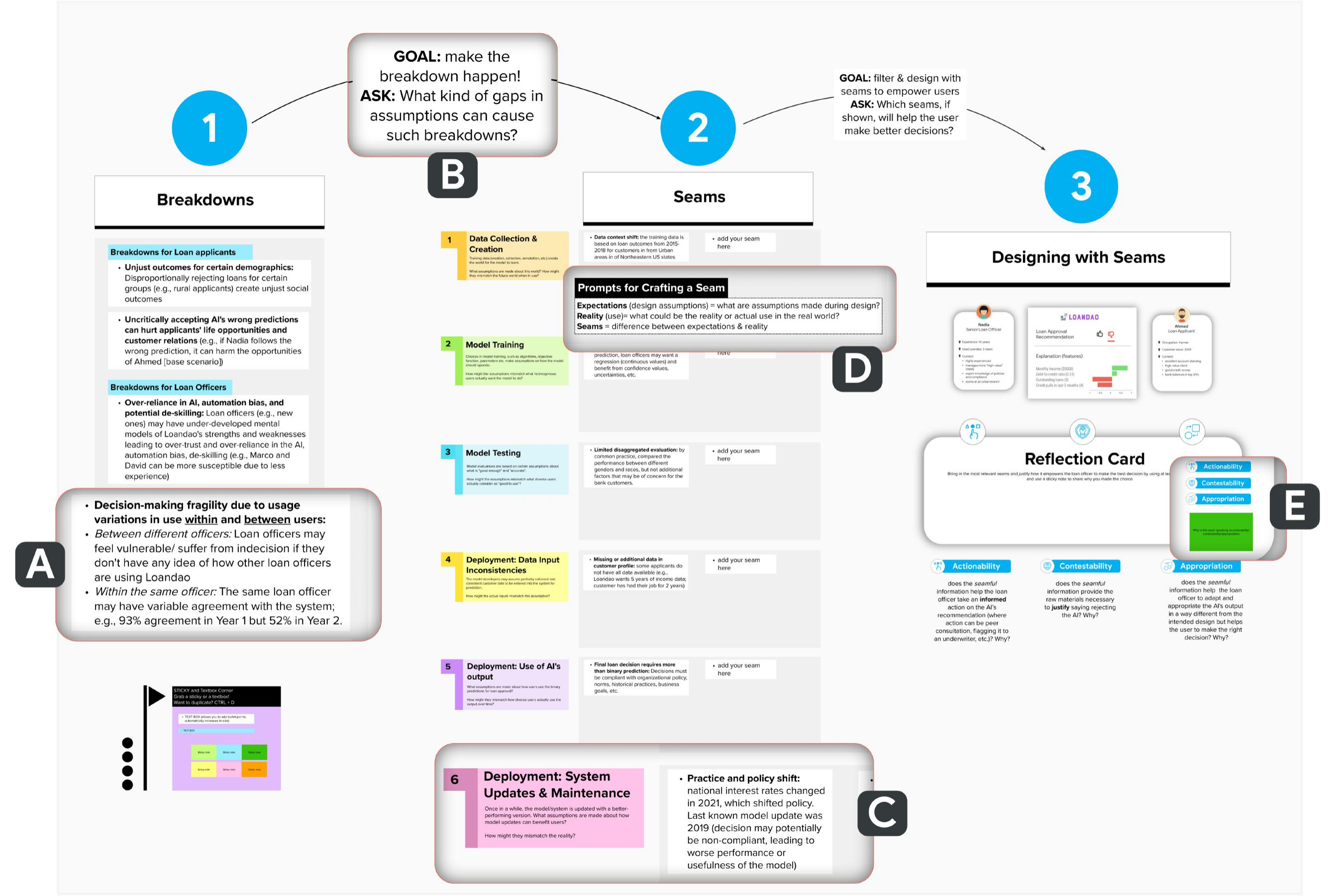}
    \vspace{-0.5em}
    \caption{The virtual whiteboard used for the seamful XAI design activity showing key features in~\cite{ehsan2022seamful} (reproduced with permission). \textbf{Area 1:} Envisioning breakdown (Step 1). Participants were provided sample breakdowns (\textbf{A}), which participants could either use directly or get inspiration for their own envisioning. \textbf{Area 2:} Anticipating \& crafting seams (Step 2). Fuiding prompts were provided (\textbf{B}) for effectively crafting the seams. Exemplary seams were shared (\textbf{C}) for each stage of the AI lifecycle framework. \textbf{Area 3:} Designing with seams (Step 3). Participants were asked to articulate their reasoning for choosing a seam and tag which user goals the selected seam (\textbf{E}) can support for augmenting user agency.
    } 
   \label{fig:mural_design_process}
    \Description[figure]{Figure 5: A screenshot of the virtual whiteboard used for the Seamful XAI design activity in the study, with zoomed-in examples. There are three main areas highlighted. Area 1: Envisioning breakdowns (this is step 1). In this study, sample breakdowns were provided(highlighted in the A box - e.g., decision making fragility due to usage variations between users). Participants could either use these directly or inspire their own envisioning. Area 2: Anticipating and crafting seams (this is step 2). Guiding prompts were provided (shown in B box - e.g., Goal is to make the breakdown happen and the ask is to identify which kinds of gaps in assumptions can cause such breakdowns). Exemplary seams were given (shown in box C - e.g., shifting practices and policies).  Area 3: Designing with seams (this is step 3). Participants were encouraged to articulate and tag which user goals the selected seam (shown in box E - e.g., actionability, contestability, or appropriation) can support for augmenting user agency. The virtual whiteboard is organized as a tripartite table with a combination of text and icons.}
    \vspace{-15pt}
\end{figure*}

The \textit{second} step is around anticipating and crafting seams, which is done in three parts. First (2A in the diagram), we ask: "what might we (as developers, designers, researchers, etc.) do to make the breakdown happen?” While this question might seem counter-intuitive, it allows us to systematically prevent breakdowns by understanding their causes. This step inverts the problem and makes it a goal directed task, which is important to generate concrete outcomes instead of open-ended problems. Next (2B), we try to anticipate the reasons for the breakdown (the seams) in the appropriate stage in the AI's lifecycle (the colored boxes numbered 1-6 in Fig.~\ref{fig:mural_design_process}). Finally (2C), we craft the seam by thinking about the gap between the ideal expectation and the reality of use.

The \textit{final} step involves using the seams generated in step 2 in a way to empower user agency and explainability. Here (3A), we ask: given our end goal, which seams do we show and which do we hide (e.g. strategic revelation and concealment)? The revealed seams (3B) should empower users through better explainability. This step of the Seamful XAI process is a major differentiator from other Responsible AI processes: unlike most processes that stop at identifying gaps, this one goes beyond. It not only uncovers the gaps but also utilizes them as avenues to support users (for more details, refer to~\cite{ehsan2022seamful}). 

A co-designing study~\cite{ehsan2022seamful} with 43 real-world AI users found three beneficial elements of Seamful XAI: 
\begin{itemize}
    \item It \textbf{enhances explainability} by helping stakeholders reveal the AI’s blind spots, highlight its fallibility, and showcase the strengths and weaknesses of the system, which can calibrate reliance in AI systems.
    \item It \textbf{augments user agency} by providing peripheral vision of the AI’s blind spots. Seamful information expands the action space of what users can do. Information in seams can convert “unknown unknowns” to “known unknowns,” which can empower users to know “where” to start an investigation.
    \item It is a resourceful way to not just reveal seams but also \textbf{anticipate and mitigate harms from AI systems}.
\end{itemize}
\vspace{-2mm}

\section{Takeaways}
\textit{We began with the provocation: With the advent of Foundation Models \& Large Language Models like ChatGPT, is “opening the black-box” still a reasonable and achievable goal for XAI? Do we need to shift our perspectives?}

\textbf{Yes}. The proverbial “black-box” of AI has evolved, and so should our expectations on how to make it explainable. As the box becomes more opaque and harder to “open,” the human side of the Human-AI assemblage remains as a fruitful space to explore. In the most extreme case, \textit{the human side may be all there is left to explore}. Even if we can open the black box it is unclear what actionable outcomes would become available.

There are four important lessons from Human-centered XAI that can inform the shift in our XAI expectations. 

\begin{enumerate}
\vspace{-2mm}
    \item First, the human-centered XAI perspective takes a pragmatic and resourceful view of explainability, especially if black boxes are expected to persist. By considering the actions afforded to the user by the explanations, HCXAI centers the focus on the human, ensuring AI augments human abilities rather than replace them.
    
    \item Second, explainability is not only achieved by looking inside the black box through mechanistic descriptions of how an algorithm works. Actionability can be achieved by exploring explainability outside and around the edges of the black box because human-centered XAI takes a more expansive view of what it means to provide insights into a black box that can afford a wider range of actions.
    
    \item Third, explicitly treating AI systems as human-AI assemblages means focusing on explainability of the assemblage, not just the AI. This widened perspective opens up avenues for not just factoring in who is interacting with the black box, but also how human teams can work together — directly or indirectly — to contextualize a dynamically changing real-world AI behavior.
    
    \item Fourth, seamful XAI turns the disadvantages and weaknesses of an AI system into advantages. The gaps between user expectations and AI capabilities are exactly the gaps that explanations address. Instead of hiding those gaps to create seamless experiences, seamful XAI leverages these gaps in an opportunistic manner to augment explainability and user agency.
\end{enumerate}

% \vspace{-2mm}
As we reload our expectations on XAI, we invite you to do what HCXAI asks us to do: centering the design and evaluation around the human. This positioning can reveal unmet needs that must be addressed while avoiding the costly mistake of building XAI systems that do not make a difference. While there have been many examples of XAI systems that have failed to have the intended impact of users, it is often the case that these tenets of HCXAI were overlooked. XAI is a relatively young field of research that has yet to find its footing, even as the landscape of black box AI systems is rapidly evolving. It is not yet time to give up hope on XAI. Instead, we invite you to adopt critical reflection and value-sensitivity into XAI research and evaluation, making it human-centered. 

\centering\textbf{\textit{Will Human-centered XAI solve all our problems? No, but it will help us ask the right questions.}}

\justifying
\begin{acks}
With our deepest gratitude, we acknowledge the time of all participants of all the studies reported here. Without their input, these projects would not have been possible. We thank reviewers for their valuable input. We also want to thank the organizations, the sites for the case studies, for their cooperation. We are grateful to members of the Human-Centered AI Lab at Georgia Tech whose continued input refined the conceptualizations presented here. 
We are indebted to Justin Weisz for his editorial feedback that helped scope the project appropriately.
This project was partially supported by the National Science Foundation under Grant No. 1928586.
\end{acks}
\bibliography{sample-base}


\begin{thebibliography}{48}

%%% the Web address.

\ifx \showCODEN    \undefined \def \showCODEN     #1{\unskip}     \fi
\ifx \showDOI      \undefined \def \showDOI       #1{#1}\fi
\ifx \showISBNx    \undefined \def \showISBNx     #1{\unskip}     \fi
\ifx \showISBNxiii \undefined \def \showISBNxiii  #1{\unskip}     \fi
\ifx \showISSN     \undefined \def \showISSN      #1{\unskip}     \fi
\ifx \showLCCN     \undefined \def \showLCCN      #1{\unskip}     \fi
\ifx \shownote     \undefined \def \shownote      #1{#1}          \fi
\ifx \showarticletitle \undefined \def \showarticletitle #1{#1}   \fi
\ifx \showURL      \undefined \def \showURL       {\relax}        \fi
% invisible to TeX
\providecommand\bibfield[2]{#2}
\providecommand\bibinfo[2]{#2}
\providecommand\natexlab[1]{#1}
\providecommand\showeprint[2][]{arXiv:#2}

\bibitem[\protect\citeauthoryear{Agre}{Agre}{1997a}]%
        {agre1997toward}
\bibfield{author}{\bibinfo{person}{P Agre}.} \bibinfo{year}{1997}\natexlab{a}.
\newblock \showarticletitle{Toward a critical technical practice: Lessons learned in trying to reform AI in Bowker}.
\newblock \bibinfo{journal}{\emph{Social science, technical systems, and cooperative work: Beyond the Great Divide}} (\bibinfo{year}{1997}).
\newblock


\bibitem[\protect\citeauthoryear{Agre}{Agre}{1997b}]%
        {agre1997computation}
\bibfield{author}{\bibinfo{person}{Philip~E Agre}.} \bibinfo{year}{1997}\natexlab{b}.
\newblock \bibinfo{booktitle}{\emph{Computation and human experience}}.
\newblock \bibinfo{publisher}{Cambridge University Press}.
\newblock


\bibitem[\protect\citeauthoryear{Alqaraawi, Schuessler, Wei{\ss}, Costanza, and Berthouze}{Alqaraawi et~al\mbox{.}}{2020}]%
        {alqaraawi2020evaluating}
\bibfield{author}{\bibinfo{person}{Ahmed Alqaraawi}, \bibinfo{person}{Martin Schuessler}, \bibinfo{person}{Philipp Wei{\ss}}, \bibinfo{person}{Enrico Costanza}, {and} \bibinfo{person}{Nadia Berthouze}.} \bibinfo{year}{2020}\natexlab{}.
\newblock \showarticletitle{Evaluating saliency map explanations for convolutional neural networks: a user study}. In \bibinfo{booktitle}{\emph{Proceedings of the 25th International Conference on Intelligent User Interfaces}}. \bibinfo{pages}{275--285}.
\newblock


\bibitem[\protect\citeauthoryear{Arrieta, D{\'\i}az-Rodr{\'\i}guez, Del~Ser, Bennetot, Tabik, Barbado, Garc{\'\i}a, Gil-L{\'o}pez, Molina, Benjamins, et~al\mbox{.}}{Arrieta et~al\mbox{.}}{2020}]%
        {arrieta2020explainable}
\bibfield{author}{\bibinfo{person}{Alejandro~Barredo Arrieta}, \bibinfo{person}{Natalia D{\'\i}az-Rodr{\'\i}guez}, \bibinfo{person}{Javier Del~Ser}, \bibinfo{person}{Adrien Bennetot}, \bibinfo{person}{Siham Tabik}, \bibinfo{person}{Alberto Barbado}, \bibinfo{person}{Salvador Garc{\'\i}a}, \bibinfo{person}{Sergio Gil-L{\'o}pez}, \bibinfo{person}{Daniel Molina}, \bibinfo{person}{Richard Benjamins}, {et~al\mbox{.}}} \bibinfo{year}{2020}\natexlab{}.
\newblock \showarticletitle{Explainable Artificial Intelligence (XAI): Concepts, taxonomies, opportunities and challenges toward responsible AI}.
\newblock \bibinfo{journal}{\emph{Information Fusion}}  \bibinfo{volume}{58} (\bibinfo{year}{2020}), \bibinfo{pages}{82--115}.
\newblock


\bibitem[\protect\citeauthoryear{Arya, Bellamy, Chen, Dhurandhar, Hind, Hoffman, Houde, Liao, Luss, Mojsilovi{\'c}, et~al\mbox{.}}{Arya et~al\mbox{.}}{2019}]%
        {arya2019one}
\bibfield{author}{\bibinfo{person}{Vijay Arya}, \bibinfo{person}{Rachel~KE Bellamy}, \bibinfo{person}{Pin-Yu Chen}, \bibinfo{person}{Amit Dhurandhar}, \bibinfo{person}{Michael Hind}, \bibinfo{person}{Samuel~C Hoffman}, \bibinfo{person}{Stephanie Houde}, \bibinfo{person}{Q~Vera Liao}, \bibinfo{person}{Ronny Luss}, \bibinfo{person}{Aleksandra Mojsilovi{\'c}}, {et~al\mbox{.}}} \bibinfo{year}{2019}\natexlab{}.
\newblock \showarticletitle{One explanation does not fit all: A toolkit and taxonomy of ai explainability techniques}.
\newblock \bibinfo{journal}{\emph{arXiv preprint arXiv:1909.03012}}  \bibinfo{volume}{abs/1909.03012} (\bibinfo{year}{2019}).
\newblock
\showeprint[arxiv]{1909.03012}
\urldef\tempurl%
\url{http://arxiv.org/abs/1909.03012}
\showURL{%
\tempurl}


\bibitem[\protect\citeauthoryear{Bhatt, Xiang, Sharma, Weller, Taly, Jia, Ghosh, Puri, Moura, and Eckersley}{Bhatt et~al\mbox{.}}{2020}]%
        {bhatt2020explainable}
\bibfield{author}{\bibinfo{person}{Umang Bhatt}, \bibinfo{person}{Alice Xiang}, \bibinfo{person}{Shubham Sharma}, \bibinfo{person}{Adrian Weller}, \bibinfo{person}{Ankur Taly}, \bibinfo{person}{Yunhan Jia}, \bibinfo{person}{Joydeep Ghosh}, \bibinfo{person}{Ruchir Puri}, \bibinfo{person}{Jos{\'e}~MF Moura}, {and} \bibinfo{person}{Peter Eckersley}.} \bibinfo{year}{2020}\natexlab{}.
\newblock \showarticletitle{Explainable machine learning in deployment}. In \bibinfo{booktitle}{\emph{Proceedings of the 2020 Conference on Fairness, Accountability, and Transparency}}. \bibinfo{pages}{648--657}.
\newblock


\bibitem[\protect\citeauthoryear{Bills, Cammarata, Mossing, Tillman, Gao, Goh, Sutskever, Leike, Wu, and Saunders}{Bills et~al\mbox{.}}{2023}]%
        {bills2023language}
\bibfield{author}{\bibinfo{person}{Steven Bills}, \bibinfo{person}{Nick Cammarata}, \bibinfo{person}{Dan Mossing}, \bibinfo{person}{Henk Tillman}, \bibinfo{person}{Leo Gao}, \bibinfo{person}{Gabriel Goh}, \bibinfo{person}{Ilya Sutskever}, \bibinfo{person}{Jan Leike}, \bibinfo{person}{Jeff Wu}, {and} \bibinfo{person}{William Saunders}.} \bibinfo{year}{2023}\natexlab{}.
\newblock \bibinfo{title}{Language models can explain neurons in language models}.
\newblock \bibinfo{howpublished}{\url{https://openaipublic.blob.core.windows.net/neuron-explainer/paper/index.html}}.
\newblock


\bibitem[\protect\citeauthoryear{Buolamwini and Gebru}{Buolamwini and Gebru}{2018}]%
        {buolamwini2018gender}
\bibfield{author}{\bibinfo{person}{Joy Buolamwini} {and} \bibinfo{person}{Timnit Gebru}.} \bibinfo{year}{2018}\natexlab{}.
\newblock \showarticletitle{Gender shades: Intersectional accuracy disparities in commercial gender classification}. In \bibinfo{booktitle}{\emph{Conference on fairness, accountability and transparency}}. PMLR, \bibinfo{pages}{77--91}.
\newblock


\bibitem[\protect\citeauthoryear{Castelvecchi}{Castelvecchi}{2016}]%
        {castelvecchi2016can}
\bibfield{author}{\bibinfo{person}{Davide Castelvecchi}.} \bibinfo{year}{2016}\natexlab{}.
\newblock \showarticletitle{Can we open the black box of AI?}
\newblock \bibinfo{journal}{\emph{Nature News}} \bibinfo{volume}{538}, \bibinfo{number}{7623} (\bibinfo{year}{2016}), \bibinfo{pages}{20}.
\newblock


\bibitem[\protect\citeauthoryear{Chalmers and MacColl}{Chalmers and MacColl}{2003}]%
        {chalmers2003seamful}
\bibfield{author}{\bibinfo{person}{Matthew Chalmers} {and} \bibinfo{person}{Ian MacColl}.} \bibinfo{year}{2003}\natexlab{}.
\newblock \showarticletitle{Seamful and seamless design in ubiquitous computing}. In \bibinfo{booktitle}{\emph{Workshop at the crossroads: The interaction of HCI and systems issues in UbiComp}}, Vol.~\bibinfo{volume}{8}.
\newblock


\bibitem[\protect\citeauthoryear{Dennett}{Dennett}{1989}]%
        {dennett1989intentional}
\bibfield{author}{\bibinfo{person}{Daniel~Clement Dennett}.} \bibinfo{year}{1989}\natexlab{}.
\newblock \bibinfo{booktitle}{\emph{The intentional stance}}.
\newblock \bibinfo{publisher}{MIT press}.
\newblock


\bibitem[\protect\citeauthoryear{Dhanorkar, Wolf, Qian, Xu, Popa, and Li}{Dhanorkar et~al\mbox{.}}{2021}]%
        {dhanorkar_who_2021}
\bibfield{author}{\bibinfo{person}{Shipi Dhanorkar}, \bibinfo{person}{Christine~T Wolf}, \bibinfo{person}{Kun Qian}, \bibinfo{person}{Anbang Xu}, \bibinfo{person}{Lucian Popa}, {and} \bibinfo{person}{Yunyao Li}.} \bibinfo{year}{2021}\natexlab{}.
\newblock \showarticletitle{Who needs to know what, when?: Broadening the Explainable AI (XAI) Design Space by Looking at Explanations Across the AI Lifecycle}. In \bibinfo{booktitle}{\emph{Designing Interactive Systems Conference 2021}}.
\newblock


\bibitem[\protect\citeauthoryear{Dourish}{Dourish}{2004}]%
        {dourish2004action}
\bibfield{author}{\bibinfo{person}{Paul Dourish}.} \bibinfo{year}{2004}\natexlab{}.
\newblock \bibinfo{booktitle}{\emph{Where the action is: the foundations of embodied interaction}}.
\newblock \bibinfo{publisher}{MIT press}.
\newblock


\bibitem[\protect\citeauthoryear{Dourish, Finlay, Sengers, and Wright}{Dourish et~al\mbox{.}}{2004}]%
        {dourish2004reflective}
\bibfield{author}{\bibinfo{person}{Paul Dourish}, \bibinfo{person}{Janet Finlay}, \bibinfo{person}{Phoebe Sengers}, {and} \bibinfo{person}{Peter Wright}.} \bibinfo{year}{2004}\natexlab{}.
\newblock \showarticletitle{Reflective HCI: Towards a critical technical practice}. In \bibinfo{booktitle}{\emph{CHI'04 extended abstracts on Human factors in computing systems}}. \bibinfo{pages}{1727--1728}.
\newblock


\bibitem[\protect\citeauthoryear{Dove, Halskov, Forlizzi, and Zimmerman}{Dove et~al\mbox{.}}{2017}]%
        {dove_ux_2017}
\bibfield{author}{\bibinfo{person}{Graham Dove}, \bibinfo{person}{Kim Halskov}, \bibinfo{person}{Jodi Forlizzi}, {and} \bibinfo{person}{John Zimmerman}.} \bibinfo{year}{2017}\natexlab{}.
\newblock \showarticletitle{{UX} {Design} {Innovation}: {Challenges} for {Working} with {Machine} {Learning} as a {Design} {Material}}.
\newblock \bibinfo{journal}{\emph{Proceedings of the 2017 CHI Conference on Human Factors in Computing Systems - CHI '17}} (\bibinfo{year}{2017}), \bibinfo{pages}{278--288}.
\newblock
\showISSN{1069-3424}
\urldef\tempurl%
\url{https://doi.org/10.1145/3025453.3025739}
\showDOI{\tempurl}


\bibitem[\protect\citeauthoryear{Ehsan, Liao, Muller, Riedl, and Weisz}{Ehsan et~al\mbox{.}}{2021a}]%
        {ehsan2021expanding}
\bibfield{author}{\bibinfo{person}{Upol Ehsan}, \bibinfo{person}{Q~Vera Liao}, \bibinfo{person}{Michael Muller}, \bibinfo{person}{Mark~O Riedl}, {and} \bibinfo{person}{Justin~D Weisz}.} \bibinfo{year}{2021}\natexlab{a}.
\newblock \showarticletitle{Expanding explainability: Towards social transparency in ai systems}. In \bibinfo{booktitle}{\emph{Proceedings of the 2021 CHI Conference on Human Factors in Computing Systems}}. \bibinfo{pages}{1--19}.
\newblock


\bibitem[\protect\citeauthoryear{Ehsan, Liao, Passi, Riedl, and Daume~III}{Ehsan et~al\mbox{.}}{2022a}]%
        {ehsan2022seamful}
\bibfield{author}{\bibinfo{person}{Upol Ehsan}, \bibinfo{person}{Q~Vera Liao}, \bibinfo{person}{Samir Passi}, \bibinfo{person}{Mark~O Riedl}, {and} \bibinfo{person}{Hal Daume~III}.} \bibinfo{year}{2022}\natexlab{a}.
\newblock \showarticletitle{Seamful XAI: Operationalizing Seamful Design in Explainable AI}.
\newblock \bibinfo{journal}{\emph{arXiv preprint arXiv:2211.06753}} (\bibinfo{year}{2022}).
\newblock


\bibitem[\protect\citeauthoryear{Ehsan, Passi, Liao, Chan, Lee, Muller, Riedl, et~al\mbox{.}}{Ehsan et~al\mbox{.}}{2021b}]%
        {ehsan2021explainable}
\bibfield{author}{\bibinfo{person}{Upol Ehsan}, \bibinfo{person}{Samir Passi}, \bibinfo{person}{Q~Vera Liao}, \bibinfo{person}{Larry Chan}, \bibinfo{person}{I Lee}, \bibinfo{person}{Michael Muller}, \bibinfo{person}{Mark~O Riedl}, {et~al\mbox{.}}} \bibinfo{year}{2021}\natexlab{b}.
\newblock \showarticletitle{The who in explainable ai: How ai background shapes perceptions of ai explanations}.
\newblock \bibinfo{journal}{\emph{arXiv preprint arXiv:2107.13509}} (\bibinfo{year}{2021}).
\newblock


\bibitem[\protect\citeauthoryear{Ehsan and Riedl}{Ehsan and Riedl}{2020}]%
        {ehsan2020human}
\bibfield{author}{\bibinfo{person}{Upol Ehsan} {and} \bibinfo{person}{Mark~O Riedl}.} \bibinfo{year}{2020}\natexlab{}.
\newblock \showarticletitle{Human-centered Explainable AI: Towards a Reflective Sociotechnical Approach}, In \bibinfo{booktitle}{International Conference on Human-Computer Interaction}.
\newblock \bibinfo{journal}{\emph{arXiv preprint arXiv:2002.01092}}, \bibinfo{pages}{449--466}.
\newblock


\bibitem[\protect\citeauthoryear{Ehsan, Saha, De~Choudhury, and Riedl}{Ehsan et~al\mbox{.}}{2023}]%
        {ehsan2023charting}
\bibfield{author}{\bibinfo{person}{Upol Ehsan}, \bibinfo{person}{Koustuv Saha}, \bibinfo{person}{Munmun De~Choudhury}, {and} \bibinfo{person}{Mark~O Riedl}.} \bibinfo{year}{2023}\natexlab{}.
\newblock \showarticletitle{Charting the Sociotechnical Gap in Explainable AI: A Framework to Address the Gap in XAI}.
\newblock \bibinfo{journal}{\emph{Proceedings of the ACM on Human-Computer Interaction}} \bibinfo{volume}{7}, \bibinfo{number}{CSCW1} (\bibinfo{year}{2023}), \bibinfo{pages}{1--32}.
\newblock


\bibitem[\protect\citeauthoryear{Ehsan, Tambwekar, Chan, Harrison, and Riedl}{Ehsan et~al\mbox{.}}{2019}]%
        {ehsan2019automated}
\bibfield{author}{\bibinfo{person}{Upol Ehsan}, \bibinfo{person}{Pradyumna Tambwekar}, \bibinfo{person}{Larry Chan}, \bibinfo{person}{Brent Harrison}, {and} \bibinfo{person}{Mark~O Riedl}.} \bibinfo{year}{2019}\natexlab{}.
\newblock \showarticletitle{Automated rationale generation: a technique for explainable AI and its effects on human perceptions}. In \bibinfo{booktitle}{\emph{Proceedings of the 24th International Conference on Intelligent User Interfaces}} (Marina del Ray, California) \emph{(\bibinfo{series}{IUI '19})}. \bibinfo{publisher}{Association for Computing Machinery}, \bibinfo{address}{New York, NY, USA}, \bibinfo{pages}{263--274}.
\newblock
\showISBNx{9781450362726}
\urldef\tempurl%
\url{https://doi.org/10.1145/3301275.3302316}
\showDOI{\tempurl}


\bibitem[\protect\citeauthoryear{Ehsan, Wintersberger, Liao, Mara, Streit, Wachter, Riener, and Riedl}{Ehsan et~al\mbox{.}}{2021c}]%
        {ehsan2021operationalizing}
\bibfield{author}{\bibinfo{person}{Upol Ehsan}, \bibinfo{person}{Philipp Wintersberger}, \bibinfo{person}{Q~Vera Liao}, \bibinfo{person}{Martina Mara}, \bibinfo{person}{Marc Streit}, \bibinfo{person}{Sandra Wachter}, \bibinfo{person}{Andreas Riener}, {and} \bibinfo{person}{Mark~O Riedl}.} \bibinfo{year}{2021}\natexlab{c}.
\newblock \showarticletitle{Operationalizing human-centered perspectives in explainable AI}. In \bibinfo{booktitle}{\emph{Extended Abstracts of the 2021 CHI Conference on Human Factors in Computing Systems}}. \bibinfo{pages}{1--6}.
\newblock


\bibitem[\protect\citeauthoryear{Ehsan, Wintersberger, Liao, Watkins, Manger, Daum{\'e}~III, Riener, and Riedl}{Ehsan et~al\mbox{.}}{2022b}]%
        {ehsan2022human}
\bibfield{author}{\bibinfo{person}{Upol Ehsan}, \bibinfo{person}{Philipp Wintersberger}, \bibinfo{person}{Q~Vera Liao}, \bibinfo{person}{Elizabeth~Anne Watkins}, \bibinfo{person}{Carina Manger}, \bibinfo{person}{Hal Daum{\'e}~III}, \bibinfo{person}{Andreas Riener}, {and} \bibinfo{person}{Mark~O Riedl}.} \bibinfo{year}{2022}\natexlab{b}.
\newblock \showarticletitle{Human-Centered Explainable AI (HCXAI): beyond opening the black-box of AI}. In \bibinfo{booktitle}{\emph{CHI Conference on Human Factors in Computing Systems Extended Abstracts}}. \bibinfo{pages}{1--7}.
\newblock


\bibitem[\protect\citeauthoryear{Fodor}{Fodor}{1975}]%
        {fodor1975language}
\bibfield{author}{\bibinfo{person}{Jerry~A Fodor}.} \bibinfo{year}{1975}\natexlab{}.
\newblock \bibinfo{booktitle}{\emph{The language of thought}}. Vol.~\bibinfo{volume}{5}.
\newblock \bibinfo{publisher}{Harvard university press}.
\newblock


\bibitem[\protect\citeauthoryear{Friedman, Kahn, and Borning}{Friedman et~al\mbox{.}}{2008}]%
        {friedman2008value}
\bibfield{author}{\bibinfo{person}{Batya Friedman}, \bibinfo{person}{Peter~H Kahn}, {and} \bibinfo{person}{Alan Borning}.} \bibinfo{year}{2008}\natexlab{}.
\newblock \showarticletitle{Value sensitive design and information systems}.
\newblock \bibinfo{journal}{\emph{The handbook of information and computer ethics}} (\bibinfo{year}{2008}), \bibinfo{pages}{69--101}.
\newblock


\bibitem[\protect\citeauthoryear{Gaver, Beaver, and Benford}{Gaver et~al\mbox{.}}{2003}]%
        {gaver2003ambiguity}
\bibfield{author}{\bibinfo{person}{William~W Gaver}, \bibinfo{person}{Jacob Beaver}, {and} \bibinfo{person}{Steve Benford}.} \bibinfo{year}{2003}\natexlab{}.
\newblock \showarticletitle{Ambiguity as a resource for design}. In \bibinfo{booktitle}{\emph{Proceedings of the SIGCHI conference on Human factors in computing systems}}. \bibinfo{pages}{233--240}.
\newblock


\bibitem[\protect\citeauthoryear{Haque, Weathington, Chudzik, and Guha}{Haque et~al\mbox{.}}{2020}]%
        {haque2020understanding}
\bibfield{author}{\bibinfo{person}{MD~Romael Haque}, \bibinfo{person}{Katherine Weathington}, \bibinfo{person}{Joseph Chudzik}, {and} \bibinfo{person}{Shion Guha}.} \bibinfo{year}{2020}\natexlab{}.
\newblock \showarticletitle{Understanding Law Enforcement and Common Peoples' Perspectives on Designing Explainable Crime Mapping Algorithms}. In \bibinfo{booktitle}{\emph{Conference Companion Publication of the 2020 on Computer Supported Cooperative Work and Social Computing}}. \bibinfo{pages}{269--273}.
\newblock


\bibitem[\protect\citeauthoryear{Joshi, Koyejo, Vijitbenjaronk, Kim, and Ghosh}{Joshi et~al\mbox{.}}{2019}]%
        {joshi2019towards}
\bibfield{author}{\bibinfo{person}{Shalmali Joshi}, \bibinfo{person}{Oluwasanmi Koyejo}, \bibinfo{person}{Warut Vijitbenjaronk}, \bibinfo{person}{Been Kim}, {and} \bibinfo{person}{Joydeep Ghosh}.} \bibinfo{year}{2019}\natexlab{}.
\newblock \showarticletitle{Towards realistic individual recourse and actionable explanations in black-box decision making systems}.
\newblock \bibinfo{journal}{\emph{arXiv preprint arXiv:1907.09615}} (\bibinfo{year}{2019}).
\newblock


\bibitem[\protect\citeauthoryear{Kaur, Nori, Jenkins, Caruana, Wallach, and Wortman~Vaughan}{Kaur et~al\mbox{.}}{2020}]%
        {kaur2020interpreting}
\bibfield{author}{\bibinfo{person}{Harmanpreet Kaur}, \bibinfo{person}{Harsha Nori}, \bibinfo{person}{Samuel Jenkins}, \bibinfo{person}{Rich Caruana}, \bibinfo{person}{Hanna Wallach}, {and} \bibinfo{person}{Jennifer Wortman~Vaughan}.} \bibinfo{year}{2020}\natexlab{}.
\newblock \showarticletitle{Interpreting Interpretability: Understanding Data Scientists' Use of Interpretability Tools for Machine Learning}. In \bibinfo{booktitle}{\emph{Proceedings of the 2020 CHI Conference on Human Factors in Computing Systems}} (Honolulu, HI, USA) \emph{(\bibinfo{series}{CHI '20})}. \bibinfo{publisher}{Association for Computing Machinery}, \bibinfo{address}{New York, NY, USA}, \bibinfo{pages}{1--14}.
\newblock
\showISBNx{9781450367080}
\urldef\tempurl%
\url{https://doi.org/10.1145/3313831.3376219}
\showDOI{\tempurl}


\bibitem[\protect\citeauthoryear{Lewis}{Lewis}{1986}]%
        {lewis1986causal}
\bibfield{author}{\bibinfo{person}{David~K Lewis}.} \bibinfo{year}{1986}\natexlab{}.
\newblock \showarticletitle{Causal explanation}.
\newblock  (\bibinfo{year}{1986}).
\newblock


\bibitem[\protect\citeauthoryear{Lewis, Perez, Piktus, Petroni, Karpukhin, Goyal, K{\"u}ttler, Lewis, Yih, Rockt{\"a}schel, et~al\mbox{.}}{Lewis et~al\mbox{.}}{2020}]%
        {lewis2020retrieval}
\bibfield{author}{\bibinfo{person}{Patrick Lewis}, \bibinfo{person}{Ethan Perez}, \bibinfo{person}{Aleksandra Piktus}, \bibinfo{person}{Fabio Petroni}, \bibinfo{person}{Vladimir Karpukhin}, \bibinfo{person}{Naman Goyal}, \bibinfo{person}{Heinrich K{\"u}ttler}, \bibinfo{person}{Mike Lewis}, \bibinfo{person}{Wen-tau Yih}, \bibinfo{person}{Tim Rockt{\"a}schel}, {et~al\mbox{.}}} \bibinfo{year}{2020}\natexlab{}.
\newblock \showarticletitle{Retrieval-augmented generation for knowledge-intensive nlp tasks}.
\newblock \bibinfo{journal}{\emph{Advances in Neural Information Processing Systems}}  \bibinfo{volume}{33} (\bibinfo{year}{2020}), \bibinfo{pages}{9459--9474}.
\newblock


\bibitem[\protect\citeauthoryear{Liao, Gruen, and Miller}{Liao et~al\mbox{.}}{2020}]%
        {liao2020questioning}
\bibfield{author}{\bibinfo{person}{Q~Vera Liao}, \bibinfo{person}{Daniel Gruen}, {and} \bibinfo{person}{Sarah Miller}.} \bibinfo{year}{2020}\natexlab{}.
\newblock \showarticletitle{Questioning the AI: Informing Design Practices for Explainable AI User Experiences}. In \bibinfo{booktitle}{\emph{Proceedings of the SIGCHI Conference on Human Factors in Computing Systems}}. ACM, \bibinfo{pages}{1--15}.
\newblock


\bibitem[\protect\citeauthoryear{Liao, Pribi{\'c}, Han, Miller, and Sow}{Liao et~al\mbox{.}}{2021}]%
        {liao2021question}
\bibfield{author}{\bibinfo{person}{Q~Vera Liao}, \bibinfo{person}{Milena Pribi{\'c}}, \bibinfo{person}{Jaesik Han}, \bibinfo{person}{Sarah Miller}, {and} \bibinfo{person}{Daby Sow}.} \bibinfo{year}{2021}\natexlab{}.
\newblock \showarticletitle{Question-driven design process for explainable ai user experiences}.
\newblock \bibinfo{journal}{\emph{arXiv preprint arXiv:2104.03483}} (\bibinfo{year}{2021}).
\newblock


\bibitem[\protect\citeauthoryear{Liao and Varshney}{Liao and Varshney}{2021}]%
        {liao2021human}
\bibfield{author}{\bibinfo{person}{Q~Vera Liao} {and} \bibinfo{person}{Kush~R Varshney}.} \bibinfo{year}{2021}\natexlab{}.
\newblock \showarticletitle{Human-centered explainable ai (xai): From algorithms to user experiences}.
\newblock \bibinfo{journal}{\emph{arXiv preprint arXiv:2110.10790}} (\bibinfo{year}{2021}).
\newblock


\bibitem[\protect\citeauthoryear{Miller}{Miller}{2019}]%
        {miller2019explanation}
\bibfield{author}{\bibinfo{person}{Tim Miller}.} \bibinfo{year}{2019}\natexlab{}.
\newblock \showarticletitle{Explanation in artificial intelligence: Insights from the social sciences}.
\newblock \bibinfo{journal}{\emph{Artificial Intelligence}}  \bibinfo{volume}{267} (\bibinfo{year}{2019}), \bibinfo{pages}{1--38}.
\newblock


\bibitem[\protect\citeauthoryear{Mohseni, Zarei, and Ragan}{Mohseni et~al\mbox{.}}{2018}]%
        {mohseni2018multidisciplinary}
\bibfield{author}{\bibinfo{person}{Sina Mohseni}, \bibinfo{person}{Niloofar Zarei}, {and} \bibinfo{person}{Eric~D Ragan}.} \bibinfo{year}{2018}\natexlab{}.
\newblock \showarticletitle{A Multidisciplinary Survey and Framework for Design and Evaluation of Explainable AI Systems}.
\newblock \bibinfo{journal}{\emph{arXiv}} (\bibinfo{year}{2018}), \bibinfo{pages}{arXiv--1811}.
\newblock
\urldef\tempurl%
\url{https://doi.org/10.1145/3387166}
\showDOI{\tempurl}
\showeprint[arxiv]{1811.11839}


\bibitem[\protect\citeauthoryear{Nott}{Nott}{2017}]%
        {nott2017explainable}
\bibfield{author}{\bibinfo{person}{George Nott}.} \bibinfo{year}{2017}\natexlab{}.
\newblock \showarticletitle{Explainable artificial intelligence: Cracking open the black box of AI}.
\newblock \bibinfo{journal}{\emph{Computer world}}  \bibinfo{volume}{4} (\bibinfo{year}{2017}).
\newblock


\bibitem[\protect\citeauthoryear{P{\'a}ez}{P{\'a}ez}{2019}]%
        {paez2019pragmatic}
\bibfield{author}{\bibinfo{person}{Andr{\'e}s P{\'a}ez}.} \bibinfo{year}{2019}\natexlab{}.
\newblock \showarticletitle{The pragmatic turn in explainable artificial intelligence (XAI)}.
\newblock \bibinfo{journal}{\emph{Minds and Machines}} \bibinfo{volume}{29}, \bibinfo{number}{3} (\bibinfo{year}{2019}), \bibinfo{pages}{441--459}.
\newblock


\bibitem[\protect\citeauthoryear{Poursabzi-Sangdeh, Goldstein, Hofman, Vaughan, and Wallach}{Poursabzi-Sangdeh et~al\mbox{.}}{2018}]%
        {poursabzi2018manipulating}
\bibfield{author}{\bibinfo{person}{Forough Poursabzi-Sangdeh}, \bibinfo{person}{Daniel~G Goldstein}, \bibinfo{person}{Jake~M Hofman}, \bibinfo{person}{Jennifer~Wortman Vaughan}, {and} \bibinfo{person}{Hanna Wallach}.} \bibinfo{year}{2018}\natexlab{}.
\newblock \showarticletitle{Manipulating and measuring model interpretability}.
\newblock \bibinfo{journal}{\emph{arXiv preprint arXiv:1802.07810}} (\bibinfo{year}{2018}).
\newblock


\bibitem[\protect\citeauthoryear{Pushkarna, Zaldivar, and Kjartansson}{Pushkarna et~al\mbox{.}}{2022}]%
        {pushkarna2022data}
\bibfield{author}{\bibinfo{person}{Mahima Pushkarna}, \bibinfo{person}{Andrew Zaldivar}, {and} \bibinfo{person}{Oddur Kjartansson}.} \bibinfo{year}{2022}\natexlab{}.
\newblock \showarticletitle{Data Cards: Purposeful and Transparent Dataset Documentation for Responsible AI}.
\newblock \bibinfo{journal}{\emph{arXiv preprint arXiv:2204.01075}} (\bibinfo{year}{2022}).
\newblock


\bibitem[\protect\citeauthoryear{Schoeffer and Kuehl}{Schoeffer and Kuehl}{2021}]%
        {schoeffer2021appropriate}
\bibfield{author}{\bibinfo{person}{Jakob Schoeffer} {and} \bibinfo{person}{Niklas Kuehl}.} \bibinfo{year}{2021}\natexlab{}.
\newblock \showarticletitle{Appropriate fairness perceptions? On the effectiveness of explanations in enabling people to assess the fairness of automated decision systems}. In \bibinfo{booktitle}{\emph{Companion Publication of the 2021 Conference on Computer Supported Cooperative Work and Social Computing}}. \bibinfo{pages}{153--157}.
\newblock


\bibitem[\protect\citeauthoryear{Sengers, Boehner, David, and Kaye}{Sengers et~al\mbox{.}}{2005}]%
        {sengers2005reflective}
\bibfield{author}{\bibinfo{person}{Phoebe Sengers}, \bibinfo{person}{Kirsten Boehner}, \bibinfo{person}{Shay David}, {and} \bibinfo{person}{Joseph'Jofish' Kaye}.} \bibinfo{year}{2005}\natexlab{}.
\newblock \showarticletitle{Reflective design}. In \bibinfo{booktitle}{\emph{Proceedings of the 4th decennial conference on Critical computing: between sense and sensibility}}. \bibinfo{pages}{49--58}.
\newblock


\bibitem[\protect\citeauthoryear{Singh, Miller, Lyons, Sonenberg, Velloso, Vetere, Howe, and Dourish}{Singh et~al\mbox{.}}{2023}]%
        {singh2023directive}
\bibfield{author}{\bibinfo{person}{Ronal Singh}, \bibinfo{person}{Tim Miller}, \bibinfo{person}{Henrietta Lyons}, \bibinfo{person}{Liz Sonenberg}, \bibinfo{person}{Eduardo Velloso}, \bibinfo{person}{Frank Vetere}, \bibinfo{person}{Piers Howe}, {and} \bibinfo{person}{Paul Dourish}.} \bibinfo{year}{2023}\natexlab{}.
\newblock \showarticletitle{Directive explanations for actionable explainability in machine learning applications}.
\newblock \bibinfo{journal}{\emph{ACM Transactions on Interactive Intelligent Systems}} (\bibinfo{year}{2023}).
\newblock


\bibitem[\protect\citeauthoryear{Stumpf, Bussone, and O’sullivan}{Stumpf et~al\mbox{.}}{2016}]%
        {stumpf2016explanations}
\bibfield{author}{\bibinfo{person}{Simone Stumpf}, \bibinfo{person}{Adrian Bussone}, {and} \bibinfo{person}{Dympna O’sullivan}.} \bibinfo{year}{2016}\natexlab{}.
\newblock \showarticletitle{Explanations considered harmful? user interactions with machine learning systems}. In \bibinfo{booktitle}{\emph{ACM SIGCHI Workshop on Human-Centered Machine Learning}}.
\newblock


\bibitem[\protect\citeauthoryear{Sun, Liao, Muller, Agarwal, Houde, Talamadupula, and Weisz}{Sun et~al\mbox{.}}{2022}]%
        {sun2022investigating}
\bibfield{author}{\bibinfo{person}{Jiao Sun}, \bibinfo{person}{Q~Vera Liao}, \bibinfo{person}{Michael Muller}, \bibinfo{person}{Mayank Agarwal}, \bibinfo{person}{Stephanie Houde}, \bibinfo{person}{Kartik Talamadupula}, {and} \bibinfo{person}{Justin~D Weisz}.} \bibinfo{year}{2022}\natexlab{}.
\newblock \showarticletitle{Investigating Explainability of Generative AI for Code through Scenario-based Design}. In \bibinfo{booktitle}{\emph{27th International Conference on Intelligent User Interfaces}}. \bibinfo{pages}{212--228}.
\newblock


\bibitem[\protect\citeauthoryear{Xie, Wiegreffe, and Riedl}{Xie et~al\mbox{.}}{2022}]%
        {xie2022calibrating}
\bibfield{author}{\bibinfo{person}{Kaige Xie}, \bibinfo{person}{Sarah Wiegreffe}, {and} \bibinfo{person}{Mark Riedl}.} \bibinfo{year}{2022}\natexlab{}.
\newblock \showarticletitle{Calibrating trust of multi-hop question answering systems with decompositional probes}.
\newblock \bibinfo{journal}{\emph{arXiv preprint arXiv:2204.07693}} (\bibinfo{year}{2022}).
\newblock


\bibitem[\protect\citeauthoryear{Yang, Steinfeld, Ros{\'e}, and Zimmerman}{Yang et~al\mbox{.}}{2020}]%
        {yang2020re}
\bibfield{author}{\bibinfo{person}{Qian Yang}, \bibinfo{person}{Aaron Steinfeld}, \bibinfo{person}{Carolyn Ros{\'e}}, {and} \bibinfo{person}{John Zimmerman}.} \bibinfo{year}{2020}\natexlab{}.
\newblock \showarticletitle{Re-examining whether, why, and how human-AI interaction is uniquely difficult to design}. In \bibinfo{booktitle}{\emph{Proc. CHI}}. \bibinfo{pages}{1--13}.
\newblock


\bibitem[\protect\citeauthoryear{Zhang, Liao, and Bellamy}{Zhang et~al\mbox{.}}{2020}]%
        {zhang2020effect}
\bibfield{author}{\bibinfo{person}{Yunfeng Zhang}, \bibinfo{person}{Q~Vera Liao}, {and} \bibinfo{person}{Rachel~KE Bellamy}.} \bibinfo{year}{2020}\natexlab{}.
\newblock \showarticletitle{Effect of Confidence and Explanation on Accuracy and Trust Calibration in AI-Assisted Decision Making}. In \bibinfo{booktitle}{\emph{Proceedings of the Conference on Fairness, Accountability, and Transparency}} (Barcelona, Spain) \emph{(\bibinfo{series}{FAT* '20})}. ACM, \bibinfo{publisher}{Association for Computing Machinery}, \bibinfo{address}{New York, NY, USA}, \bibinfo{pages}{295–305}.
\newblock
\showISBNx{9781450369367}
\urldef\tempurl%
\url{https://doi.org/10.1145/3351095.3372852}
\showDOI{\tempurl}


\end{thebibliography}
\bibliographystyle{ACM-Reference-Format}

\end{document}